\documentclass[aps,prb,twocolumn,showpacs,amssymb,amsmath,superscriptaddress,floatfix]{revtex4}
\usepackage{graphics,epsfig}

\begin{document}

\title{Frequency resonance in Josephson-junction arrays under strong driving}
\author{Jong Soo \surname{Lim}}
\affiliation{Department of Physics, Seoul National University,
Seoul 151-747, Korea}
\author{M.Y. \surname{Choi}}
\affiliation{Department of Physics, Seoul National University,
Seoul 151-747, Korea}
\affiliation{Korea Institute for Advanced Study, Seoul 130-012, Korea}
\author{Beom Jun \surname{Kim}}
\affiliation{Department of Molecular Science and Technology, Ajou University,
Suwon 442-749, Korea}

\begin{abstract}
We study resonance behavior of a two-dimensional fully frustrated Josephson-junction 
array driven by high alternating currents.
The signal-to-noise ratio (SNR) is examined as the frequency of the driving current
is varied; revealed is a certain frequency range where the SNR
is enhanced.  Such resonance behavior is explained by considering the dynamic
order parameter at zero temperature.  We also compute the work, which
corresponds to the Ohmic dissipation of the energy due to external
currents, and discuss its possibility as a measure of stochastic resonance.
\end{abstract}


\maketitle

%
Conventionally, the stochastic resonance (SR) stands for the phenomenon that a
weak signal can be amplified and optimized by the assistance of noise.
The SR, in general requiring three basic ingredients, i.e., an energetic
activation barrier or a form of threshold, a coherence input (such as
periodic driving), and a source of noise, has been observed in a variety of
systems.~\cite{gammaitoni}  It has also been investigated in the regime of
strong periodic driving.~\cite{pankratov,mahato} In particular,
Pankratov~\cite{pankratov} considered fluctuation dynamics of a Brownian
particle in a quartic potential subject to strong periodic driving, and
studied the SR behavior reflected in the signal-to-noise ratio (SNR) at
given driving amplitude, as the frequency and the noise intensity are varied.
The observed enhancement of the SR behavior in a certain range of the
driving frequency, which may be called {\em frequency resonance} for convenience, 
was explained in terms of the very weak dependence of
the mean transition time on the noise intensity in that frequency range.
From the practical viewpoint, such characteristic behavior is important
in systems with many degrees of freedom, which constitute many devices 
operating in the strong driving regime.
Among interesting systems with many degrees of freedom is
the two-dimensional (2D) fully frustrated Josephson-junction array (FFJJA),
which grants a direct experimental realization and provides rich 
dynamics.~\cite{mon,marconi,gsj}
Recent study includes dynamic transitions as well as SR in FFJJAs driven by
uniform alternating currents.~\cite{jslim}
However, responses to strong driving with the possibility of 
frequency resonance have not been examined
in a system with many degrees of freedom including the FFJJA.

This work examines the FFJJA driven by high alternating currents,
with attention paid to the frequency resonance behavior. 
In order to describe the dynamic responses,
we consider the staggered magnetization, the average of which defines the dynamic
order parameter, and compute its power spectrum. 
As manifested by the corresponding SNR, the system indeed exhibits frequency 
resonance: the SR-like behavior as the driving frequency is varied.  
It is discussed in view of the variation of the zero-temperature
states with the driving frequency.  In addition, we also 
obtain the work performed by the driving currents and explore the possibility 
as a new measure of the resonance behavior.~\cite{iwai}

%
We begin with the equations of motion for the phase angles \{$\phi_i$\} of
the superconducting order parameters on the grains forming an
$L \times L$ square lattice.
Within the resistively-shunted-junction model
under the fluctuating twist boundary conditions (FTBC),~\cite{bjk} we have
\begin{equation}
{\sum_j}' \left[ \frac{d{\widetilde{\phi}}_{ij}}{dt} +
\sin({\widetilde{\phi}}_{ij} - {\bf r}_{ij}\cdot{\bf \Delta}) + \eta_{ij} \right] = 0,
\label{eq:rsj1}
\end{equation}
where the primed summation runs over the nearest neighbors of grain $i$ and
the abbreviations
${\widetilde{\phi}}_{ij} \equiv \phi_i - \phi_j - A_{ij}$ and
${\bf r}_{ij} \equiv {\bf r}_i - {\bf r}_j$ with ${\bf r}_i = (x_i, y_i)$
denoting the position of grain $i$ have been used.
The thermal noise current $\eta_{ij}$ satisfies
$\langle \eta_{ij}(t) \eta_{kl}(t') \rangle =
2T\delta(t-t')(\delta_{ik}\delta_{jl} - \delta_{il}\delta_{jk})$ at temperature $T$
in units of $J/k_B$ with the Josephson coupling strength $J$.
Note that ${\bf r}_{ij}$ for nearest neighboring grains is a unit vector
since the lattice constant has been set equal to unity throughout this paper.
We have also expressed the energy and the time in units of
$\hbar I_c /2e$ and $\hbar /2eRI_c$, respectively,
with the critical current $I_c \equiv 2eJ/\hbar$ and the shunt resistance $R$. 
While in the FTBC the phase variables observe the periodic boundary conditions, 
the dynamics of the twist variables ${\bf \Delta} \equiv (\Delta_x, \Delta_y)$
is governed by the equations
\begin{align}
& \frac{d\Delta_x}{dt}
  = \frac{1}{L^2} \sum_{{\langle ij \rangle}_x}\sin({\widetilde{\phi}}_{ij} -\Delta_x)
    + \eta_{\Delta_x} - I_0\sin(\Omega t)  \nonumber \\
& \frac{d\Delta_y}{dt}
  = \frac{1}{L^2} \sum_{{\langle ij \rangle}_y}\sin({\widetilde{\phi}}_{ij} - \Delta_y)
    + \eta_{\Delta_y},
\label{eq:rsj2}
\end{align}
where $\sum_{\left\langle ij \right\rangle_a}$ denotes the summation over all nearest
neighboring pairs in the $a \,(= x, y)$ direction, $\eta_{\Delta_a}$ satisfies
$\left\langle \eta_{\Delta_a}(t+\tau)\eta_{\Delta_{a'}}(t) \right\rangle
= (2T/L^2)\delta(\tau)\delta_{a,a'}$,
and the oscillating current $I_0 \sin(\Omega t)$ is injected in the $x$ direction.
In the Landau gauge, the bond angle $A_{ij}$, given by the line integral of
the vector potential, takes the values
\begin{equation*}
A_{ij} =
\begin{cases}
0       & \textrm{for ${\bf r}_j = {\bf r}_i + {\bf \hat{x}}$} ,  \\
\pi x_i & \textrm{for ${\bf r}_j = {\bf r}_i + {\bf \hat{y}}$} .
\end{cases}
\end{equation*}

For the study of the dynamic behavior,
associated with the ${\rm Z}_2$ symmetry in the FFJJA,
it is convenient to consider the chirality
\begin{equation}
C({\bf R},t) \equiv {\rm sgn}
\left[ \sum_{\bf P} \sin \left( \widetilde{\phi}_{ij}(t)
      - {\bf r}_{ij}\cdot {\bf \Delta}(t) \right) \right]
\end{equation}
and the staggered magnetization
\begin{equation}
m(t) \equiv \frac{1}{L^2} \sum_{\bf R} (-1)^{x_i + y_i} C({\bf R},t),
\end{equation}
where $ \sum_{\bf P}$ denotes the directional plaquette summation of links
around the dual lattice site
${\bf R} \equiv {\bf r}_i + (1/2)(\hat{\bf x}+\hat{\bf y})$.
The dynamic order parameter is then given by the staggered magnetization averaged over
one period of the oscillating current:\cite{jslim}
\begin{equation}
Q \equiv \frac{\Omega}{2\pi} \left| \oint m(t) dt \right|.
\end{equation}
Unlike in Ref.~\onlinecite{jslim}, we keep the driving amplitude fixed
at the value larger than the critical one (mostly $I_0 =1.2$) 
and probe the behavior as the frequency $\Omega$ is varied.

In the numerical calculation, the set of the equations of motion in
Eqs.~(\ref{eq:rsj1}) and (\ref{eq:rsj2}) are integrated via the modified
Euler method with time steps of size $\Delta t = 0.04$.
We have varied the step size, only to find no essential difference.
Typically, data have been taken after the initial 500 driving periods,
which have turned out to be long enough to achieve proper stationarity.
In addition, 500 independent runs have been performed, over which the average
has been taken.

\begin{figure}
\epsfig{width=0.4\textwidth,file=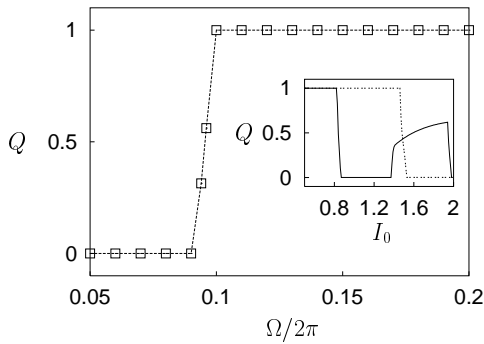}
\vspace{0.3cm}
\caption{Dynamic order parameter $Q$ versus frequency $\Omega$ 
at zero temperature, for driving amplitude $I_0 = 1.2$ 
in the system of size $L = 16$.
Inset : $Q$ versus $I_0$ for  $\Omega/2\pi = 0.06$ (solid line) and $\Omega/2\pi = 0.12$
(dotted line).}
\label{fig:Q}
\end{figure}


The behavior of the dynamic order parameter $Q$ with the driving 
amplitude $I_0$ was investigated, revealing that at zero temperature
the dynamically ordered state ($Q > 0$) and the disordered one ($Q=0$)
appear alternatively as $I_0$ is raised
(see the inset of Fig.~\ref{fig:Q}).~\cite{jslim} 
Here we examine how $Q$ changes as the driving frequency $\Omega$ is varied for
fixed $I_0$ and display the result in Fig.~\ref{fig:Q}. 
Such behavior of $Q$ with $\Omega$ can be understood as follows: 
At low driving frequencies, the external current changes with time 
more slowly than the intrinsic dynamics and thus the system has enough 
time to follow the external driving.  Consequently, the two ground-state 
configurations of the chirality alternate in time, leading to the 
oscillation of the staggered magnetization $m(t)$.
From the definition of the dynamic order parameter, this in turn 
results in $Q=0$. On the other hand, for larger values of
$\Omega$, the system cannot change the staggered magnetization in
time since the driving changes too fast in comparison with
the intrinsic time scale, and there follows $Q = 1$.

%

\begin{figure}
\centering
\epsfig{width=0.4\textwidth,file=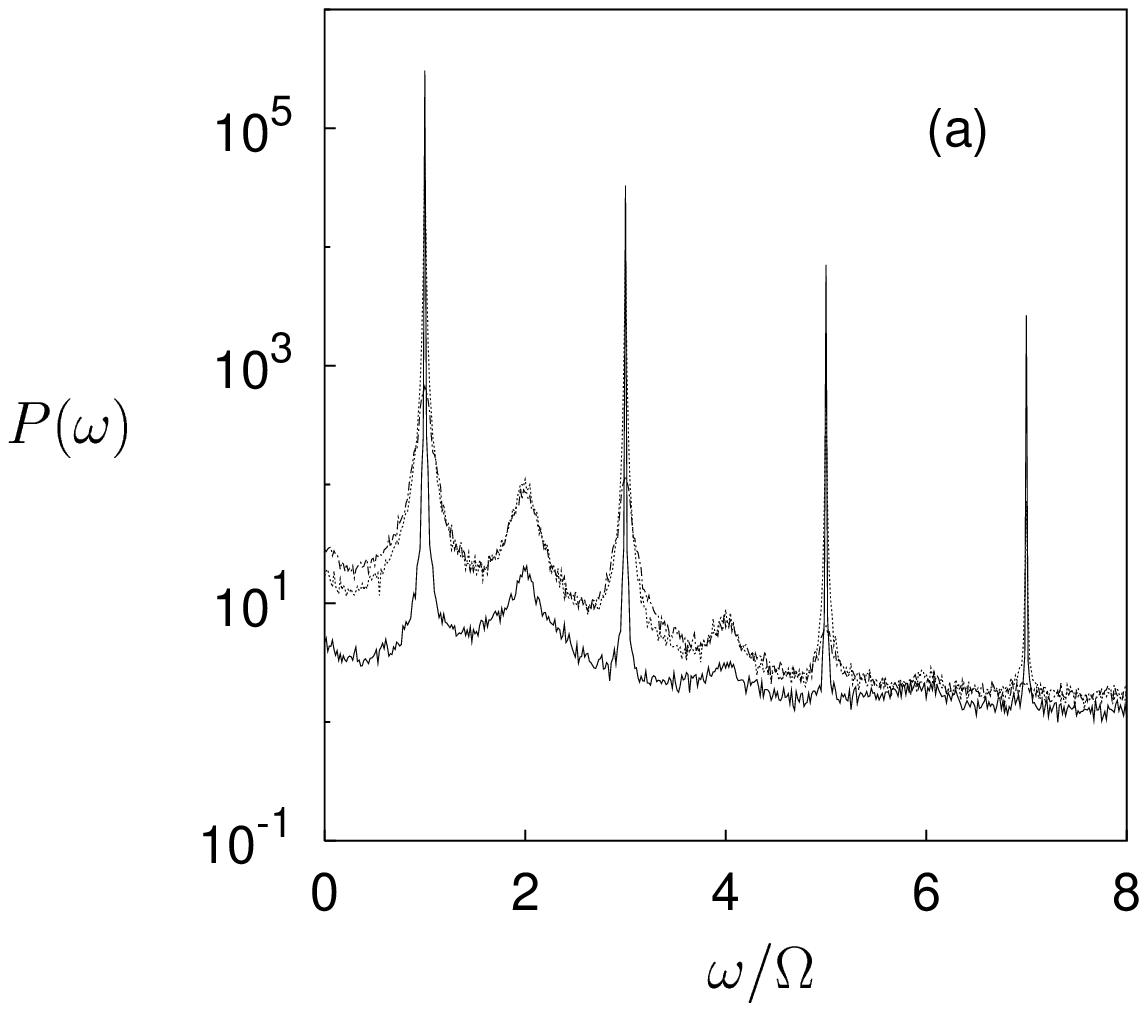}
\epsfig{width=0.4\textwidth,file=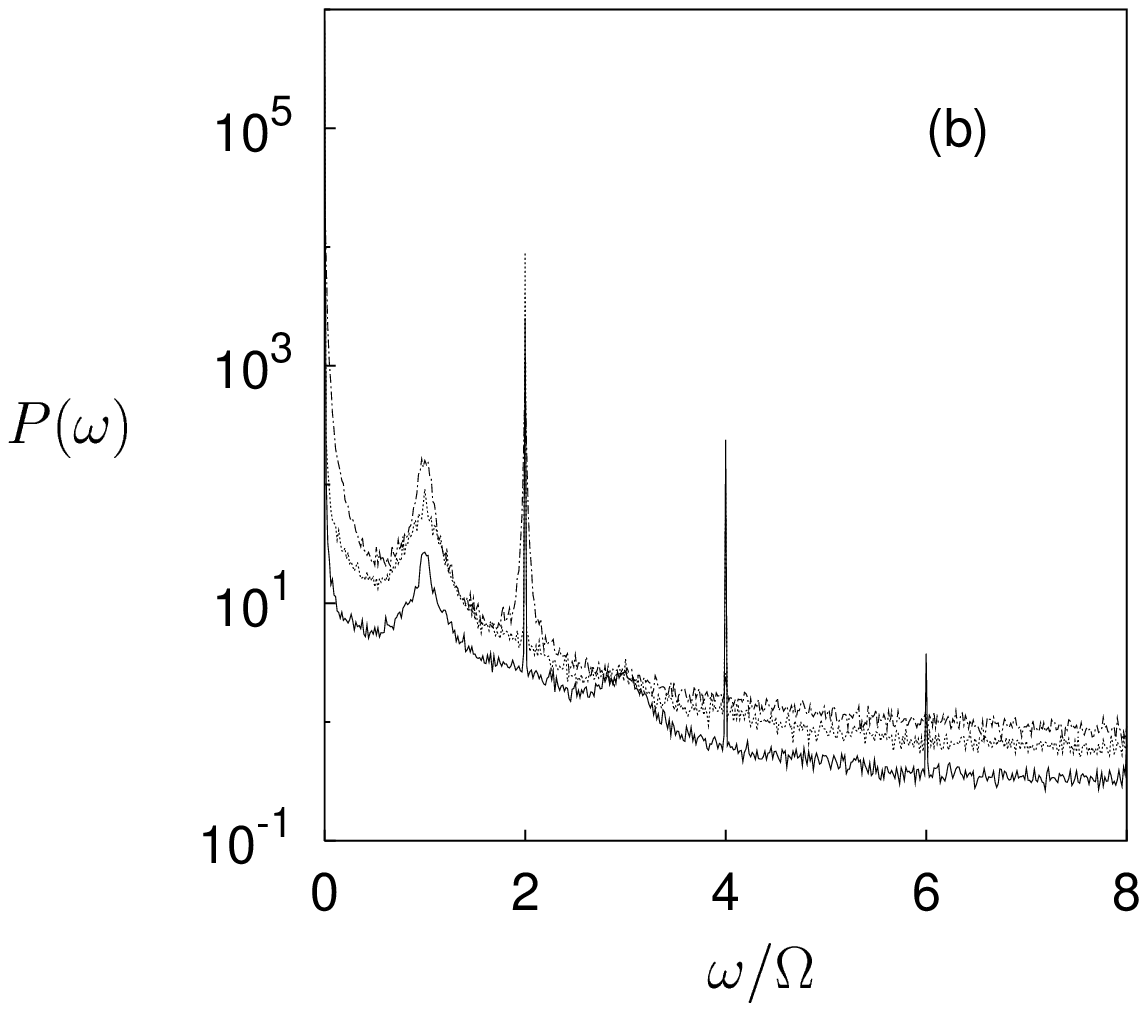}
\vspace{0.3cm}
\caption
{Power spectrum $P(\omega)$ of the staggered magnetization in the system 
of size $L = 16$, driven with the amplitude $I_0 = 1.2$ and the frequency 
$\Omega/2\pi = $ (a) $0.06$ and (b) $0.12$.
The temperature is given by $T = 0.1, 0.2$, and $0.3$ from below.}
\label{fig:power}
\end{figure}

We next investigate the resonance behavior, focusing
on the power spectrum of the staggered magnetization and its SNR defined to be
\begin{equation}
{\cal R} = 10 \log_{10} \left[ \frac{S}{N} \right].
\end{equation}
The signal $S$ represents the peak power intensity at the driving
frequency $\Omega$ while the background noise level $N$
is estimated by the average power spectrum around the signal peak.

\begin{figure}
\centering
\epsfig{width=0.4\textwidth,file=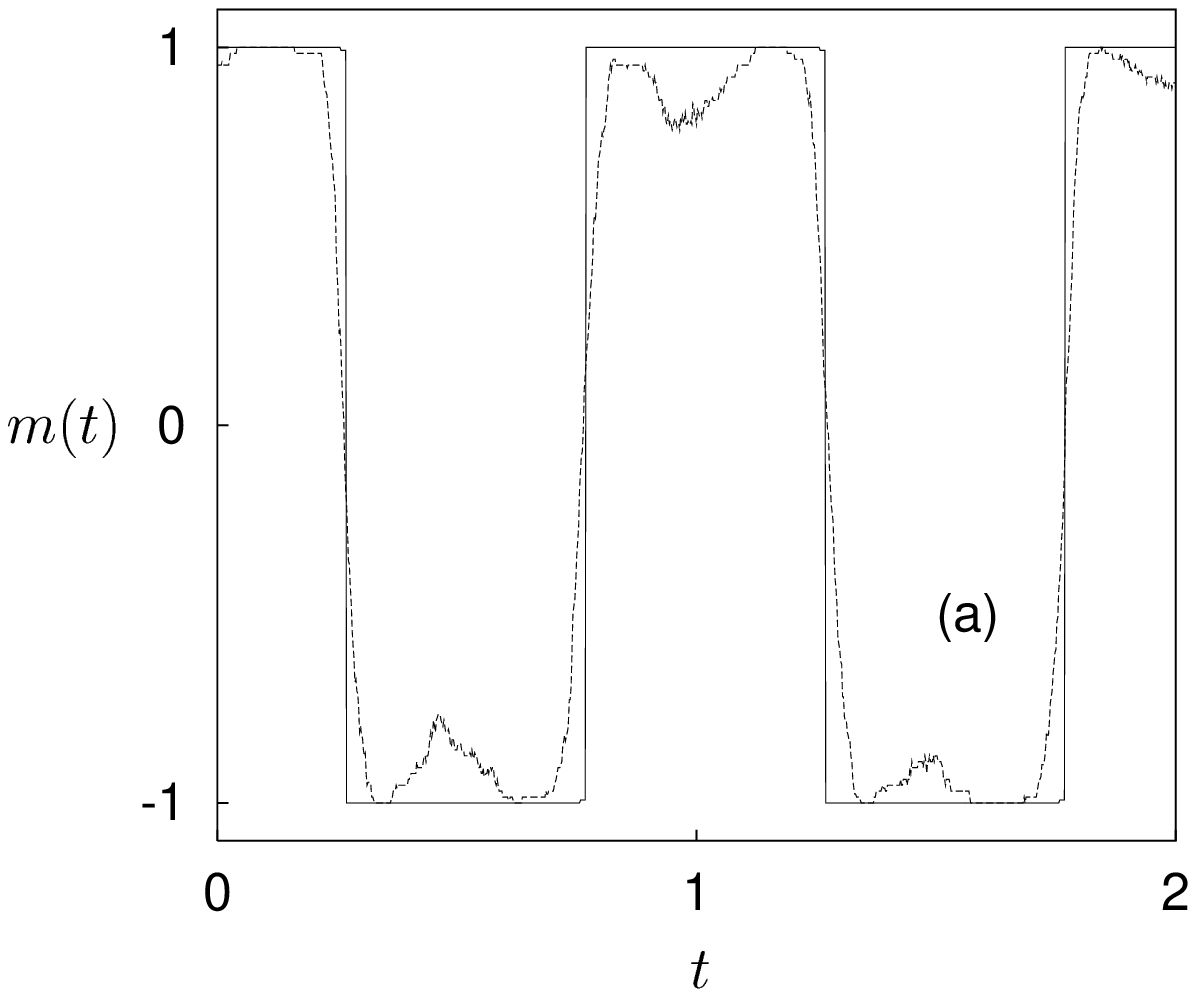}
\epsfig{width=0.4\textwidth,file=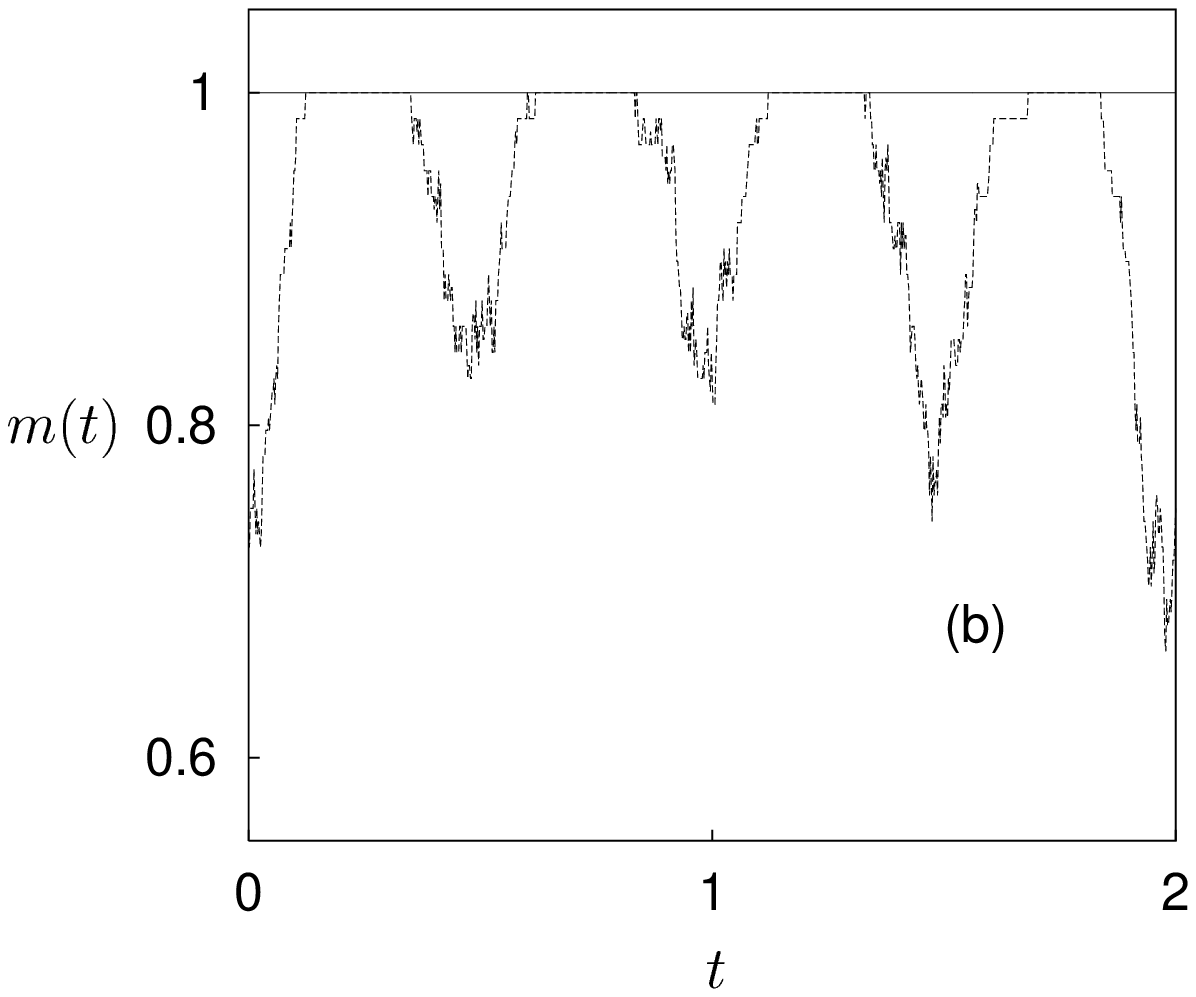}
\vspace{0.3cm}
\caption
{Time evolution of the staggered magnetization $m(t)$ for driving amplitude 
$I_0 = 1.2$ and frequency $\Omega/2\pi =$ (a) $0.06$ and (b) $0.12$. 
The temperature is given by $T = 0$ (solid line) and $0.1$ (dotted line);
time $t$ in the horizontal axis is expressed in units of the period $2\pi/\Omega$. 
}
\label{fig:mt}
\end{figure}

In Fig.~\ref{fig:power}, we display typical behavior of the power spectrum
$P(\omega)$ at low temperatures $T=0.1, 0.2$, and $0.3$ in the presence of
the driving current with amplitude $I_0 = 1.2$ and frequency 
$\Omega/2\pi =$ (a) 0.06 and (b) 0.12.  Note the substantial
difference in the behavior of $P(\omega)$ at $\Omega/2\pi =0.06$ and
at $\Omega/2\pi =0.12$: In the former $P(\omega$) exhibits pronounced peaks 
at odd harmonics ($\omega = \Omega, 3\Omega, 5\Omega, \cdots$), 
whereas even harmonics instead of odd ones are pronounced in the latter.
Somewhat similar features were also observed in the opposite case that 
the amplitude is varied with the frequency kept fixed.~\cite{jslim}

\begin{figure}
\centering
\epsfig{width=0.4\textwidth,file=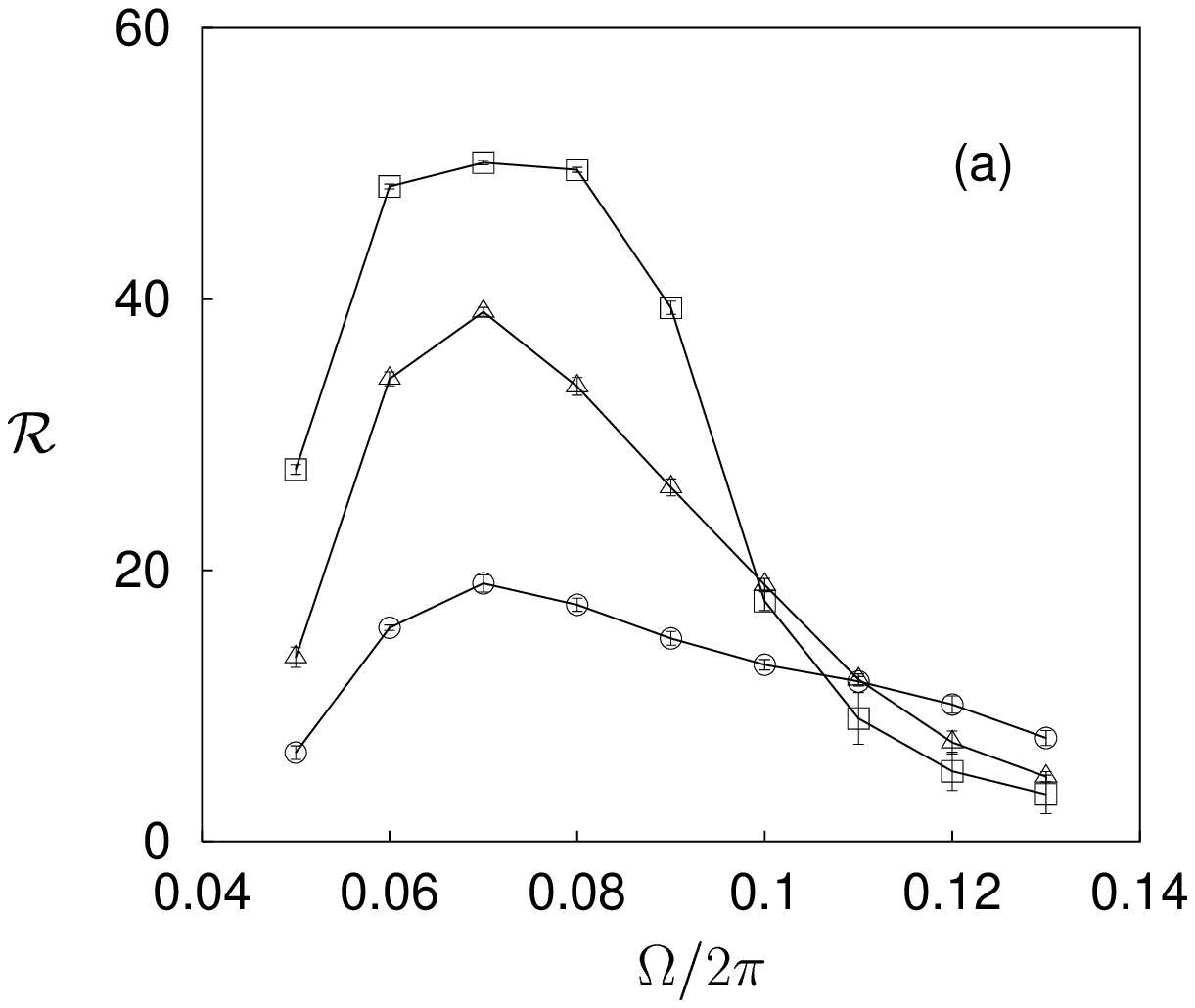}
\epsfig{width=0.4\textwidth,file=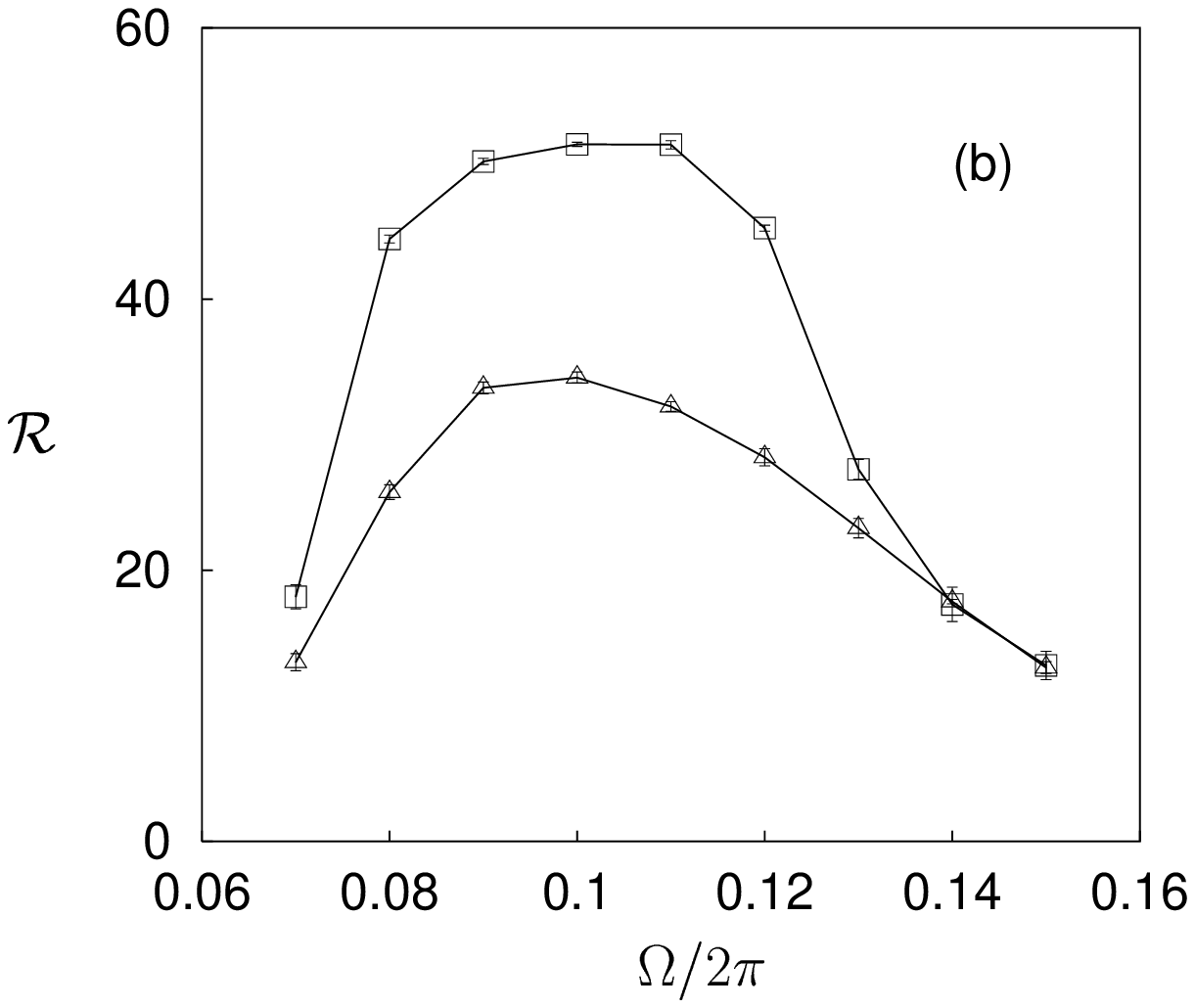}
\vspace{0.3cm}
\caption
{Behavior of the signal-to-noise ratio ${\cal R}$ with the driving frequency $\Omega$,
measured in the system of size $L=16$ at the driving amplitude
$I_0 = $ (a) $1.2$ and (b) $1.6$. Included in (a) are data at
temperature $T=0.1, 0.2$, and $0.3$ while in (b) $T= 0.1$ and $0.2$ 
(from top to bottom).
Error bars are less than the size of symbols and lines are merely guides to eyes.}
\label{fig:snr}
\end{figure}

This interesting difference between slow and fast driving is manifested by
the zero-temperature behavior of the staggered magnetization,
shown Fig.~\ref{fig:mt}:
At low driving frequencies such as $\Omega/2\pi = 0.06$, the staggered 
magnetization exhibits periodic oscillations in perfect accord with 
the external driving at zero temperature, generating pronounced odd harmonics. 
As thermal noise is introduced here, the sharp peaks at odd harmonics are 
gradually suppressed while broad peaks at even harmonics emerge. 
On the other hand, at high frequencies the system cannot follow the external 
driving so that even harmonics are prominent. 
The peak at $\omega = 0$ for $\Omega/2\pi = 0.12$ reflects the nonzero value 
of $Q$, i.e., the nonzero time-average value of $m(t)$.

Figure~\ref{fig:snr} shows the SNR versus the driving frequency at the 
amplitude $I_0 = $ (a) $1.2$ and (b) $1.6$.  
As the frequency $\Omega$ is increased, the SNR first increases, 
reaches its maximum at frequency $\Omega_{R}$, 
which we call the resonance frequency, and then decreases with further increase 
of $\Omega$. 
Such characteristic behavior can again be explained by considering
the zero temperature behavior:
At low driving frequencies, we have $Q = 0$ since $m(t)$
oscillates between $1$ and $-1$ at zero temperature. 
As the frequency $\Omega$ is increased, the driving current changes with time 
rapidly, making the system reside more in the out-of-equilibrium state.
Since in equilibrium vortices stay at positions 
where the lattice pinning potential is minimum,\cite{lattice_potential}
the increase of $\Omega$ can be interpreted as the effective reduction in the 
barrier due to the lattice pinning potential.  
Accordingly, the escape from the potential well occurs more easily, 
leading to the increase of the SNR, as shown in Fig.~\ref{fig:snr}.
Beyond the resonance frequency $\Omega_{R}$, however, further increase of $\Omega$ 
suppresses the SNR since there is no enough time to escape from the potential well.
This argument is also supported by Fig.~\ref{fig:snr}(b),
where for comparison the behavior of the SNR at the amplitude $I_0 = 1.6$ is shown. 
It is observed that the resonance frequency $\Omega_{R}$ shifts toward higher values, 
which is easily understood in view of that the escape from the well occurs at a
higher frequency for a larger value of $I_0$. 
Note also that the maximum value of the SNR decreases with the temperature. 

%
It was proposed that the work done by the external driving
can be used as a quantitative measure to detect SR; indeed
the work as a function of temperature has been shown
to have maximum near the SR temperature $T_{SR}$~\cite{iwai}.
Such resonant behavior was explained by the increase of the work as 
the resonant response is enhanced. 
In this section, we investigate the work $W$ done by the
external driving current in the FFJJA:
\begin{eqnarray}
W  &=& -\frac{I_0 \Omega }{n} \int_{t_0}^{n{\cal{T}}+t_0} 
       \left\langle \sum_y (\phi_{0,y} - \phi_{N,y}) \cos(\Omega t) \right\rangle dt 
                 \nonumber \\
   &=& \frac{I_0}{n} \int_{t_0}^{n{\cal{T}}+t_0} \left\langle V \sin(\Omega t) 
       \right\rangle dt ,
\label{eq:work}
\end{eqnarray}
where ${\cal{T}} \equiv 2\pi/\Omega$ is the period of the driving current,
$\langle \cdots \rangle$ denotes the ensemble average, and the time-average
is taken during $n$ periods of driving current.
To obtain the second line, we have integrated by parts and used the Josephson relation 
$V = (\hbar/2e) (d\phi/dt)$.  The same equation can also be derived from
the Ohmic dissipation [with the power $V(t)I(t)$ at time $t$]
averaged over $n$ periods.

\begin{figure}
\epsfig{width=0.4\textwidth,file=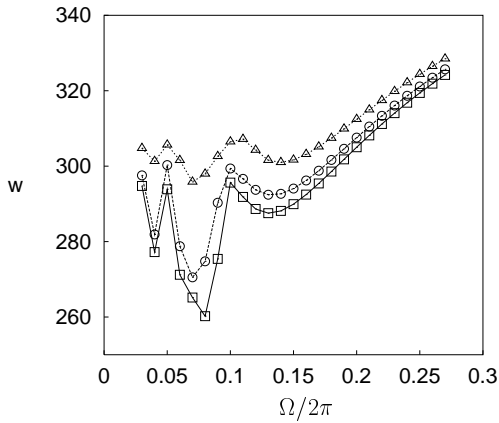}
\caption{Work (per site) done by the driving current, 
as a function of the frequency $\Omega$, in the system of size $L=8$.  
The driving amplitude is given by $I_0 = 1.2$
and the temperature $T=0.05, 0.10$, and $0.20$ from below.}
\label{fig:work}
\end{figure}

In Fig.~\ref{fig:work}, we display how the work per site $w\equiv L^{-2}W$ 
changes with the frequency $\Omega$. 
Unexpectedly, the work does not exhibit any resonant behavior, in sharp
contrast with the result in Ref.~\onlinecite{iwai}. 
We presume that this discrepancy results from the following fact:
While the work in Ref.~\onlinecite{iwai} is related to the position of 
the Brownian particle, the voltage $V$ giving the work in Eq.~(\ref{eq:work}) 
corresponds to the velocity of a vortex rather than the position. 
It is thus not likely that the work serves as an appropriate measure to 
quantify frequency resonance in our system.

%
In summary, we have studied dynamic properties of the 2D FFJJA driven uniformly by
alternating currents, with attention paid to the frequency resonance phenomena
in the strong driving regime.
For this purpose, we have computed the power spectrum of 
the staggered magnetization and the corresponding SNR. 
It has been shown that there exists a certain frequency range 
where the SNR is enhanced. 
Such behavior may be understood in view of the variation of the zero-temperature 
states with the driving frequency. 
Direct experimental observation of these dynamic properties is presumably 
possible with the state-of-art vortex imaging technique.\cite{tonomura} 
In addition, we have calculated the work done by the driving current, 
and probed its possibility as a new quantitative measure of resonance. 
It has been found that the work does not display any resonant behavior, 
suggesting inadequacy as a measure of resonance. 

%
This work was supported in part by the BK21 Program, the SKOREA Program (M.Y.C.), 
and the KOSEF Grant No. R14-2002-062-01000-0 (B.J.K.).

\end{document}